A Queueing Model of Patient Flow for Stroke Networks to Estimate Acute Stroke Transfer Capacity

Hojun Choi[1], Ohad Perry[1], Jane L. Holl[2], Shyam Prabhakaran[2]

## ABSTRACT

### BACKGROUND

Most acute stroke (AS) patients in the United States are initially evaluated at a primary stroke center (PSC) and a significant proportion requires transfer to a comprehensive stroke center (CSC) for advanced treatment. A CSC typically accepts patients from multiple PSCs in its network, leading to capacity limits. This study uses a simple queueing model to estimate impacts on CSC capacity due to transfers from PSCs.

### METHODS

The model assumes that the number of AS patients arriving at each PSC, proportion of AS patients transferred, and length of stay in the CSC Neurologic Intensive Care Unit (Neuro-ICU) by type of AS are random, while the transfer rates of ischemic and hemorrhagic AS patients are control variables. The main outcome measure is the "overflow" probability, namely, the probability of a CSC not having capacity (unavailability of a Neuro-ICU bed) to accept a transfer. Data simulations of the model, using a base case and an expanded case, were performed to illustrate the effects of changing key parameters, such as transfer rates from PSCs and CSC Neuro-ICU capacity on overflow capacity.

### RESULTS

Data simulations of the model using a base case show that an increase of a PSC's ischemic stroke transfer rate from 15% to 55% raises the overflow probability from 30.62% to 36.13%. Further simulations of the expanded case show that to maintain an *a priori* CSC overflow probability of 30.62% when adding a PSC with a AS transfer rate of 15% to the network, other PSCs would need to decrease their transfer rate by 12.5% or the CSC Neuro-ICU would need to add 2 beds.

[1] Northwestern University
[2] University of Chicago

**DISCUSSION**

A queuing model can be used to estimate the effects of change in the size of a PSC-CSC network, change in AS transfer rates, or change in number of CSC Neuro-ICU beds of a CSC on its capacity on the overflow probability in the CSC.

## INTRODUCTION

Most patients in the United States (US) with signs and symptoms suggestive of acute stroke (AS) are evaluated initially in the Emergency Department (ED) of a hospital that is either a non-stroke or primary stroke center (PSC). About 10-15% of AS patients, however, require transfer to a comprehensive stroke center (CSC) for additional evidence-based advanced treatment, such as endovascular therapy (EVT)[1-5] that cannot be delivered at a PSC.[6] Over the past decade, due to the time-sensitive nature of AS treatment, numerous initiatives have focused on speeding up, for example, the time needed for initial evaluation of AS (e.g., direct to CT protocol)[7,8] or time to administration of alteplase (e.g., door-to-needle time).[9,10] However, less work has focused on streamlining and speeding up an essential step of the AS transfer process, "securing" a bed in the Neurologic Intensive Care Unit (Neuro-ICU) at the receiving CSC.[9,11] Indeed, most CSCs accept transfers from a network of PSCs and experience fluctuations in capacity to accept AS transfers. Furthermore, a CSC Neuro-ICU also receives requests for direct admission of AS patients from its own ED. Increasing the number of PSCs in a CSC network increases the number of transfer requests, which impacts CSC Neuro-ICU bed availability.[12] Delay in "securing" CSC acceptance of an AS transfer from a PSC can lead to delay in timely receipt of an advanced treatment such as EVT.

This study evaluated a simple queueing model for CSCs to pre-emptively estimate the effect on its capacity to accept AS transfers when adding another PSC to a network, changing AS transfer rates as a result of criteria modifications in an AS transfer protocol, or changing the number of CSC Neuro-ICU beds.

## METHODS

Queueing theory is a branch of engineering and mathematics concerned with the modeling and analyses of randomly evolving systems that process a factor called "work." In a typical scenario, "work" arrives in a system and "queues" prior to receiving service if there are no available servers, and then is processed by the server before it departs; for example, customers calling a call center, "queuing" or being on hold for the next available agent, and then being served by an agent. In healthcare, queuing models have been used to study the flow of patients arriving in EDs, with some patients then being hospitalized. The *dynamics* of arrivals into the system are random because the number of calls or number of patients arriving are neither scheduled nor known, and service times or the length of queuing are also unknown at the outset.

In this study, we assessed a queueing network to model the patient-flow of AS patients from PSCs to a CSC. The model considers a network of one or more PSCs that transfer AS patients to one CSC. Beds in the CSC's Neuro-ICU are considered as the "servers." The CSC receives requests for a bed in the Neuro-

ICU for an AS patient who is transferred from PSCs in its network or directed from its own ED. Two different types of "work" arrive at PSCs: patients with acute ischemic stroke (AIS) and patients with hemorrhagic stroke (HS). While nearly all HS patients, initially evaluated at a PSC, require transfer to a CSC, only a subset of AIS patients typically require transfer to a CSC. Furthermore, although AIS accounts for 85% of AS, this proportion can vary in different populations and affect overall AS transfer rates.[13]

In the model, AS patients arrive at a PSC in a random manner, but at a known rate per day, which can be estimated from a PSC's historical data. After ED evaluation, a decision is made to transfer the patient to the CSC or not, often based on an AS transfer protocol. While nearly all HS patients require transfer, the proportion of AIS patients who need to be transferred varies. Transfer errors, due to either unnecessary transfers or "missed" transfers of patients who should have been transferred, can be estimated, using either PSC transfer rate data. Further, one can vary the transfer rate to test the effect of a change in transfer rate on CSC Neuro-ICU bed availability.

The length of stay (LOS) of AS patients in the Neuro-ICU also affects bed availability. LOS varies by type of AS, with HS patients having substantially longer LOS.[14] It is significant that the actual LOS of any individual patient is random, but its distribution can be estimated from data using statistical analysis.

To illustrate the potential of the model, data simulations were performed to estimate the effects of change in the transfer rates of AS patients from network PSCs and of increase in the number of network PSCs on the "overflow" probability, or the probability that a CSC *does not* have a Neuro-ICU bed available. The discrete-event simulation included 100 replications, each of which was run over 10,000 days (time units), and statistics over the first 2,000 days were discarded to ignore transient effects.

In the data simulations, we considered a *base* case of a network of one CSC and three (3) PSCs for simulations of the model. The parameters used in each scenario are shown in Table 1. In the base case, an average of 2 AS patients, 15% of whom are HS patients and 85% AIS patients, arrive in the ED of each PSC for evaluation; an average of 3 AS patients per day, with the same proportion of HS and AIS patients as above, arrive in the CSC's ED. For the base case, we assumed that each PSC transfers 95% of HS patients and 15% of AIS patients to the CSC. The mean Neuro-ICU LOS for HS and AIS patients was 7 and 3 days, respectively, regardless of whether the patient was transferred from a PSC or admitted directly from the CSC's ED. In the base case, the Neuro-ICU had 15 beds. We varied the transfer rate of AIS patients from one PSC to the CSC, ranging from 15% to 35% and then to 55%, with all else being constant, to estimate the effect on the overflow probability.

We then performed data simulations on the model for an *expanded* case of a network of one CSC and four (4) PSCs, with the same PSC AS patient arrival rates (for both AIS and HS), transfer rates, and Neuro-ICU LOS as the base case, and estimated the effects of changes in the parameters on the overflow probability of the CSC Neuro-ICU.

**RESULTS**

The data simulations of the model for the base case (Figure 1A) shows the effect under different PSC transfer rates of 15%, 35%, and 55%, while also accounting for AS patients admitted directly from the CSC ED, on the overflow probability. A 15% transfer rate results in at least 12 beds being occupied for 84% of the time, which rises to 87% and 89%, respectively, when the transfer rate increases to 35% and 55%. Figure 1B shows the CSC overflow probabilities by different transfer rates (30.62 ± 0.09%; 33.44 ± 0.09%; and 36.13 ± 0.08%). Table 2 summarizes the results of varying transfer rates on the CSC overflow probability and proportion of time that at least 12 beds are occupied. For transfer rates of 35% and 55%, the overflow probability could be maintained at 30.62% (the overflow probability at a 15% transfer rate) by increasing the number of available CSC Neuro-ICU beds by 1 (total of 16) or 2 (total of 17), respectively.

Table 3 shows the results of the data simulations of the model for the *expanded case*, assuming the same base case parameters (arrival rates of both types of AS patients). The proportion of time that at least 12 beds are occupied rises from 84.10% to 89.74%, from 86.57% to 91.41%, and from 88.64% to 92.53%, compared to the base case, as the transfer rate changes from 15% to 35% and 55%, respectively; and CSC overflow probabilities increase by approximately 6-7%, relative to the base case. In order to maintain the CSC bed availability similar to the base case with an expanded network, a CSC would need to decrease the AS transfer rate from the other PSCs or increase the number of CSC Neuro-ICU beds.

Figure 2 shows overflow probabilities for the (i) base case, (ii) expanded case, (iii) expanded case with transfer rates from other PSCs reduced to 2.5% (reduced by 12.5% from 15%), and (iv) expanded case with the addition of 2 CSC Neuro-ICU beds. For example, to maintain an overflow probability of the base case at 30.62%, the expanded case would require a CSC to either reduce the other PSCs' transfer rates by 12.5% or add two more beds to the CSC.

**DISCUSSION**

The transfer of AS patients within a network of multiple PSCs to a CSC is a highly dynamic process affected by multiple parameters that vary, including the number of Neuro-ICU beds available at a CSC, average Neuro-ICU LOS, the number of AS patients being evaluated in the ED ("arrival rate") of each PSC or the CSC, and PSC transfer rates by type of AS (AIS or HS). This paper proposes a queueing model to estimate the effect of these parameters on the CSC's overflow probability. The model accounts for varying transfer rates from a PSC in the network, such as a change in the AS transfer protocol or change in a PSC's AS arrival rate. It can also consider adding or removing a PSC from the CSC network and changes in CSC Neuro-ICU LOS or in the number of CSC Neuro-ICU beds available. In the simulation, the average PSC and CSC arrival rates, average PSC transfer rates, average CSC LOS, and the total number of CSC Neuro-ICU beds available are needed to estimate the CSC overflow probability. Conversely, the model can be used to set an *a priori* CSC overflow probability and determine acceptable PSC transfer rates, number of PSCs in the network, and total number of CSC Neuro-ICU beds needed to maintain the target overflow probability. (The codes for simulation can be found in this link: https://github.com/hjtree0825/stroke_network_ctmc_simulations.)

As stroke systems of care become increasingly organized and CSC hub and PSC spoke networks expand, we anticipate that that CSC bed availability will become a determining factor in the CSC's ability to accept PSC requests for AS patient transfer. Stroke programs, hospital administrators, and public policy makers will need to forecast the effects of hospital transfer protocols for stroke patients in a given network or networks of hospitals based on the likelihood of CSC bed availability. For example, if a CSC hospital is considering a merger or acquisition of another PSC hospital that will be transferring AS patients, assessment of potential Neuro-ICU bed availability and probability of overflow would help determine the transfer criteria, anticipated transfer rate from the PSC, and staffing required at the CSC.

Estimate of an overflow probability also has implications for patient care and quality of care. For instance, insufficient Neuro-ICU bed availability results in AS patients not being primarily admitted to the Neuro-ICU but to other units, that may not have nurses or physicians with neurologic expertise, thereby impacting their quality of care. Alternatively, staffing policies for nursing and physician teams across ICUs would be required to accommodate more transfers at the CSC. Finally, Neuro-ICU admissions of non-AS patients, such as elective neurosurgical admissions and ED admissions, would also be affected by increasing transfers. Thus, having a simulation tool to assess these impacts will be critical to hospital and program planning in advance of adding new hospitals to networks or changing transfer criteria.

There are several limitations to the model. We assume that AS patient "arrival" at PSCs is independent of everything else and that average CSC Neuro-ICU LOS is independently and identically distributed by AS type. However, we cannot necessarily justify these assumptions. For instance, CSC Neuro-ICU LOS may be correlated with other factors, such as the time of discharge. In addition, the model does not consider temporal dependencies, such as hourly, let alone daily, variations of the factors. While we could potentially address this issue by dividing the day into multiple time blocks (such that the parameters can be considered static within each interval), the model cannot capture other factors including ED crowding and disruptions of care that may occur, since our simulation estimates long-run averages of the outcomes. Therefore, future work that considers time-varying parameters or fitting of the distributions of arrival times and LOS could be informative. Similarly, effects of other types of AS transfer protocol-related changes could also be further examined (e.g., decrease in available Neuro-ICU beds, such as during the early phase of the COVID-19 pandemic) or major changes in a PSC-CSC stroke network due to consolidation of health systems.

**CONCLUSION**

With approximately 10-15% of AS patients, initially evaluated at a PSC, needing transfer to a CSC, it could be helpful for a CSC to have an efficient tool to estimate the effect of changes in AS transfer rates, in the number of PSCs in a CSC network, and in the total number of available Neuro-ICU beds on the overflow probability of the CSC. We employed a simple queueing model, a finite-state continuous-time Markov chain, for a PSC-CSC AS network with the CSC's overflow probability as the main outcome of interest. We provide a *base* case example to illustrate the performance outcome variation with changes in transfer protocols, CSC network composition (number of PSCs), or number of CSC Neuro-ICU beds. The model also allows for setting *a priori* the overflow probability and assessing acceptable transfer rates, number of PSCs, or total number of Neuro-ICU beds needed to maintain the probability of Neuro-ICU bed availability.

| Table 1. Base Case and Modified Parameters of the Patient-Flow Model | | | | | | | | | |
|---|---|---|---|---|---|---|---|---|---|
| | # Evaluated/day | | | | Average LOS (days) | | # Neuro-ICU Beds | Transfer Rate | |
| | PSC | | CSC | | | | | | |
| | HS | AIS | HS | AIS | HS | AIS | | HS | AIS |
| Base Case | 0.3 | 1.7 | 0.45 | 2.55 | 7 | 3 | 15 | 0.95 | 0.15 |
| Case 1: +1 PSC | 0.3 | 1.7 | 0.45 | 2.55 | | | 15 | 0.95 | 0.15 |
| Case 2: +1 PSC ↓Transfer rate | 0.3 | 1.7 | 0.45 | 2.55 | | | 15 | 0.95 | 0.025 |
| Case 3: +1 PSC ↑CSC beds | 0.3 | 1.7 | 0.45 | 2.55 | | | 17 | 0.95 | 0.15 |

| Table 2. Impact of Varied Transfer Rates on CSC "Overflow" Probability and Expanded Bed Capacity | | | |
|---|---|---|---|
| Transfer rate (%) | 15.0 | 35.0 | 55.0 |
| Overflow probability (%) | 30.62 ± 0.09 | 33.44 ± 0.09 | 36.13 ± 0.08 |
| Proportion of time that at least 12 beds are occupied (%) | 84.10 | 86.52 | 88.64 |
| Additional CSC bed capacity | 0 | 1 | 2 |
| Overflow probability with added beds (%) | Unchanged | 29.60 ± 0.09 | 28.75 ± 0.09 |
| Proportion of time that at least 12 beds are occupied with added bed capacity (%) | Unchanged | 90.45 | 94.45 |

| Table 3. Performance of Base Case and Expanded Network By Transfer Rate | | | |
|---|---|---|---|
| Transfer rate (%) | 15.0 | 35.0 | 55.0 |
| Base case: Overflow probability (%) | 30.62 ± 0.09 | 33.44 ± 0.09 | 36.13 ± 0.08 |
| Base case: Proportion of time that at least 12 beds are occupied (%) | 84.10 | 86.52 | 88.64 |
| Expanded network: Overflow probability (%) | 37.64 ± 0.07 | 40.10 ± 0.10 | 42.15 ± 0.08 |
| Expanded network: Proportion of time that at least 12 beds are occupied (%) | 89.74 | 91.41 | 92.53 |

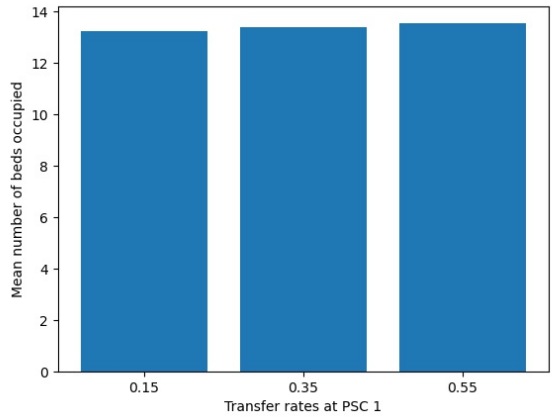

Figure 1A. Mean number of beds occupied at CSC Neuro-ICU by Transfer Rate

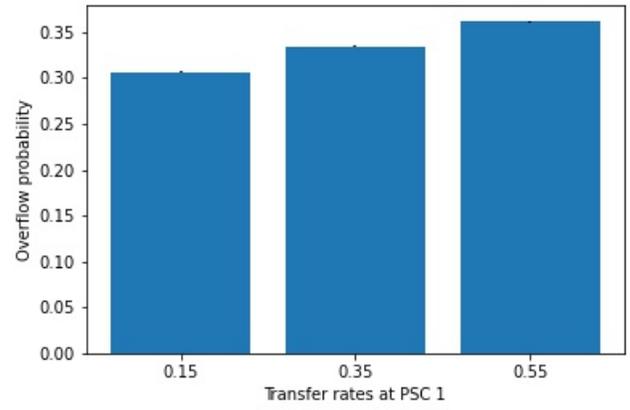

Figure 1B. CSC Overflow Probability by Transfer Rate

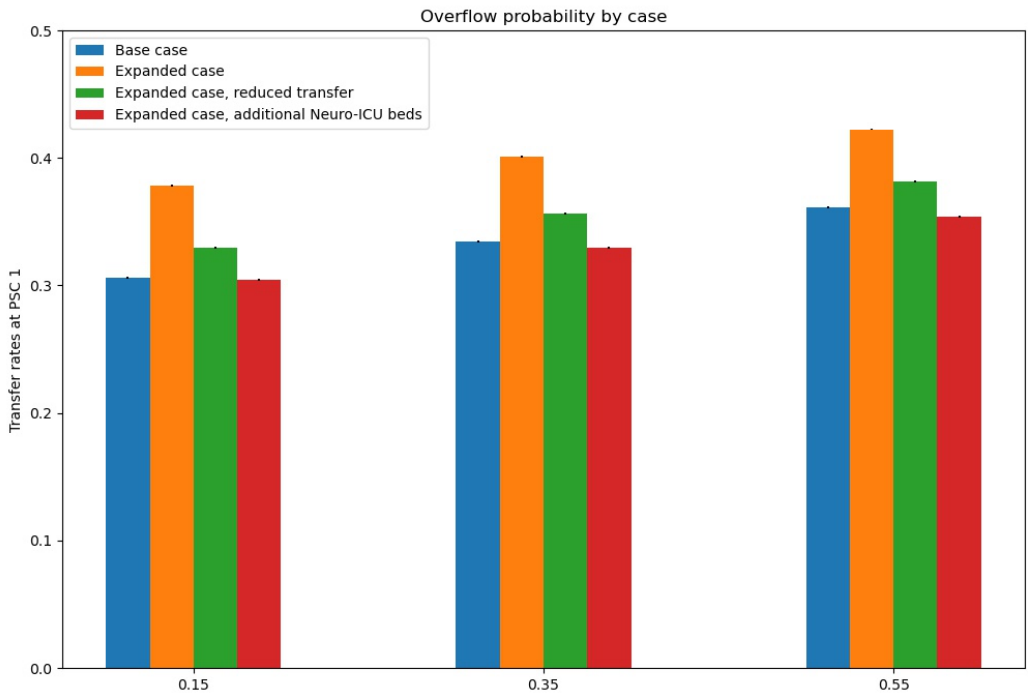

Figure 2. Overflow Probabilities by Case and by Transfer Rates at One PSC